\documentclass[twocolumn,notitlepage,prl,superscriptaddress,longbibliography]{revtex4-2}
\usepackage{amsfonts}
\usepackage{textcomp}
\usepackage{times}
\usepackage{graphicx}
\usepackage{float}
\usepackage{latexsym,amsmath,amssymb,bm,euscript}
\usepackage{color}
\usepackage{subfigure}
\usepackage{epstopdf}
\usepackage[colorlinks,
linkcolor=blue,
anchorcolor=blue,
citecolor=blue,
urlcolor=blue,]{hyperref}
\usepackage{hyperref}
\usepackage{soul}
\usepackage[normalem]{ulem}
\usepackage{mathrsfs}
\usepackage{amsmath}
\usepackage{xspace}
\usepackage{natbib}
\usepackage{ulem}
\usepackage{etoolbox}
\usepackage{tikz}
\usepackage{array}

\begin{document}
\title{Unifying Strain-driven and Pressure-driven Superconductivity in La$ _{3} $Ni$ _{2} $O$ _{7} $: Suppressed charge/spin density waves and enhanced interlayer coupling}
	
		\author{Xin-Wei Yi}
	\affiliation{School of Physical Sciences, University of Chinese Academy of Sciences, Beijing 100049, China}
	
	\author{Wei Li}
	\affiliation{Institute of Theoretical Physics, Chinese Academy of Sciences, Beijing 100190, China}
	
	\author{Jing-Yang You}
	\email{phyjyy@buaa.edu.cng}
	\affiliation{Peng Huanwu Collaborative Center for Research and Education, Beihang University, Beijing 100191,China,}
	
	\author{Bo Gu}
	\email{gubo@ucas.ac.cn}	
	\affiliation{School of Physical Sciences, University of Chinese Academy of Sciences, Beijing 100049, China} \affiliation{Kavli Institute for Theoretical Sciences, University of Chinese Academy of Sciences, Beijing 100190, China}
	
	\author{Gang Su}
	\email{gsu@ucas.ac.cn}
	\affiliation{School of Physical Sciences, University of Chinese Academy of Sciences, Beijing 100049, China} \affiliation{Institute of Theoretical Physics, Chinese Academy of Sciences, Beijing 100190, China} \affiliation{Kavli Institute for Theoretical Sciences, University of Chinese Academy of Sciences, Beijing 100190, China}
	
	\begin{abstract}

Recent strain-stabilized superconductivity at ambient pressure in La$_3$Ni$_2$O$_{7}$ films opens new avenues for nickelates research, in parallel with its pressure-induced counterpart. Using density functional theory calculations, we elucidate the critical factors bridging strain- and pressure-driven superconductivity in La$_3$Ni$_2$O$_{7}$ by comprehensively analyzing structural, electronic, magnetic, and density wave characteristics. Consistent with recent scanning transmission electron microscopy observations, we find an $I4/mmm$ structural transition at $-0.9\%$ strain, preceding superconductivity onset. Electronic analysis shows compressive strain lowers Ni-$d_{z^2}$ orbital energy levels, while interfacial Sr diffusion effectively reconstructs the $d_{z^2}$ pockets, quantitatively matching angle-resolved photoemission spectroscopy experiments. The interlayer antiferromagnetic coupling $J_\perp$ under pressure or strain closely tracks experimental superconducting $T_c$ variation. The dome-shaped pressure dependence and monotonic strain dependence of $J_\perp$ mainly arise from modulations in the apical oxygen $p_z$ energy levels. Moreover, compressive strain suppresses both charge density waves (CDW) and spin density waves (SDW) instabilities analogous to pressure effects, with SDW vanishing concurrently with the structural transition and CDW disappearing at $\sim-3.3\%$ strain. Our results indicate that suppressed density waves and enhanced $J_\perp$ are crucial for both strain- and pressure-driven superconductivity. Accordingly, we propose several candidate substrates capable of achieving greater compressive strain, thereby potentially increasing $T_c$.

	\end{abstract}
	
	\maketitle
	

$\textit{Introduction.}$---The recent discovery of high-temperature superconductivity in pressurized La$_{3}$Ni$_{2}$O$_{7}$ (superconducting $T_c\approx80$ K) has revitalized interest in nickelates as platforms for unconventional superconductivity \cite{RN1300, 2307.14819, RN1310, RN1515, RN1417, RN1384, 2311.12361}. The 4d$^{7.5}$ electronic configuration (Ni$^{2.5+}$) of La$ _{3} $Ni$ _{2} $O$ _{7} $ lies between Cu${^{2+}}$ in cuprates and Fe${^{2+}}$ in iron-based superconductors \cite{RN1300, RN1389, RN1308, 2412.18343, RN1333, 2309.17279,RN902, RN1535}. Its bilayer structure and multi-orbital nature suggest that established mechanisms for other unconventional superconductors may not directly apply to this system, presenting both challenges and opportunities \cite{RN1532, RN1361}. Significant differences between the ambient-pressure and high-pressure phases have been identified, such as structural transitions, suppression of density waves, metallization of $d_{z^2}$ bands, increased orbital hybridization, strengthened interlayer coupling, and enhanced electronic correlations \cite{RN1300, RN1473, RN1417, RN1571, RN1641, RN1596, RN1400, RN1308, RN1534, RN1424, RN1475, RN1321, 2311.12769, RN1327}. Despite extensive theoretical investigations, consensus on the key factors driving superconductivity remains elusive \cite{RN1534, RN1400, RN1399, RN1323, RN1321, RN1350, 2306.07275, RN1324, RN1326, RN1329, RN1331, RN1332, RN1334, RN1335, RN1451, RN1337, RN1397, RN1340, RN1342, 2308.11195, RN1351, RN1372, 2309.15095, 2309.17279, RN1390, RN1430, RN1424, 2311.03349, RN1453, RN1455, 2311.09970, RN1475, 2311.12769, RN1519, RN1520, RN1521, RN1534, RN1553, RN1475, RN1334, RN1967}.

Experimental progress under pressure has improved structural purity and superconducting volume fraction, but high-pressure conditions hinder full characterization and practical applications \cite{RN1458, RN1473, RN1417, 2501.14584}. As a pressure alternative, epitaxial thin-film growth has led to ambient-pressure superconductivity in compressively strained La$_{3}$Ni$_{2}$O$_{7}$ films (strain $\sim -2\%$) with $T_{\text{c}}^{\text{zero}} \approx 2$ K confirmed by zero-resistance measurements \cite{RN1943}. Optimization of growth conditions raised $T_c$ onsets above 48 K, with both zero resistance and Meissner diamagnetism observed $T_{\text{c}}^{\text{zero}}\approx 9$ K \cite{RN2003}. Mirroring bulk La$_{2}$PrNi$_{2}$O$_{7}$ \cite{RN1458}, isovalent Pr substitution further enhanced film quality, yielding $T_{\text{c}}^{\text{zero}}$ above 30 K and critical current densities of $\sim 10^4$ A$\cdot$cm$^{-2}$ at 1.4 K \cite{2501.08022}. Thin-film samples enable detailed probing of superconducting states. Recent angle-resolved photoemission spectroscopy (ARPES) measurements on La$_{2.85}$Pr$_{0.15}$Ni$_{2}$O$_{7}$ films revealed a diffuse $\gamma$ pocket from bonding $d_{z^2}$ orbitals replicating pressure effects in bulk systems, alongside a pseudogap above $T_c$,  \cite{2501.09255,2502.17831}. Conversely, another ARPES investigation on strained La$_{2}$PrNi$_{2}$O$_{7}$ films indicate bonding $d_{z^2}$ orbitals stays $\sim70$ meV below the Fermi level (E$_F$) \cite{2504.16372}. Despite separate progress in bulk and thin-film La$_{3}$Ni$_{2}$O$_{7}$ studies, direct comparisons between pressure- and strain-driven superconductivity remain limited. Clarifying their parallels and distinctions is crucial for identifying key mechanisms underlying superconductivity in La$_{3}$Ni$_{2}$O$_{7}$.

In this letter, we systematically compare the structural, electronic, magnetic, and density wave properties of La$_3$Ni$_2$O$_{7}$ under pressure or strain. A phase transition to a high-symmetry phase is found at $-0.9\%$ strain, preceding the superconducting onset near $-2\%$ strain. Electronic structure and tight-binding analyses demonstrate that compressive strain lowers the Ni-$d_{z^2}$ orbital energy, while Sr doping reconstructs the $\gamma$ pockets. Magnetic calculations further reveal that the interlayer antiferromagnetic coupling $ J_\perp $ under pressure or strain exhibits striking consistency with the experimental variation of $ T_c $. Notably, under compressive strain, spin density wave (SDW) vanishes concurrently with the structural transition, while charge density wave (CDW) is gradually suppressed and disappears at $\sim-3.3\%$ strain. These findings highlight the suppressed density waves and enhanced $ J_\perp $ are critical for superconductivity, suggesting that further increasing compressive strain could potentially elevate $ T_c $.


\begin{figure}[!t]
	\centering
	\includegraphics[scale=0.3,angle=0]{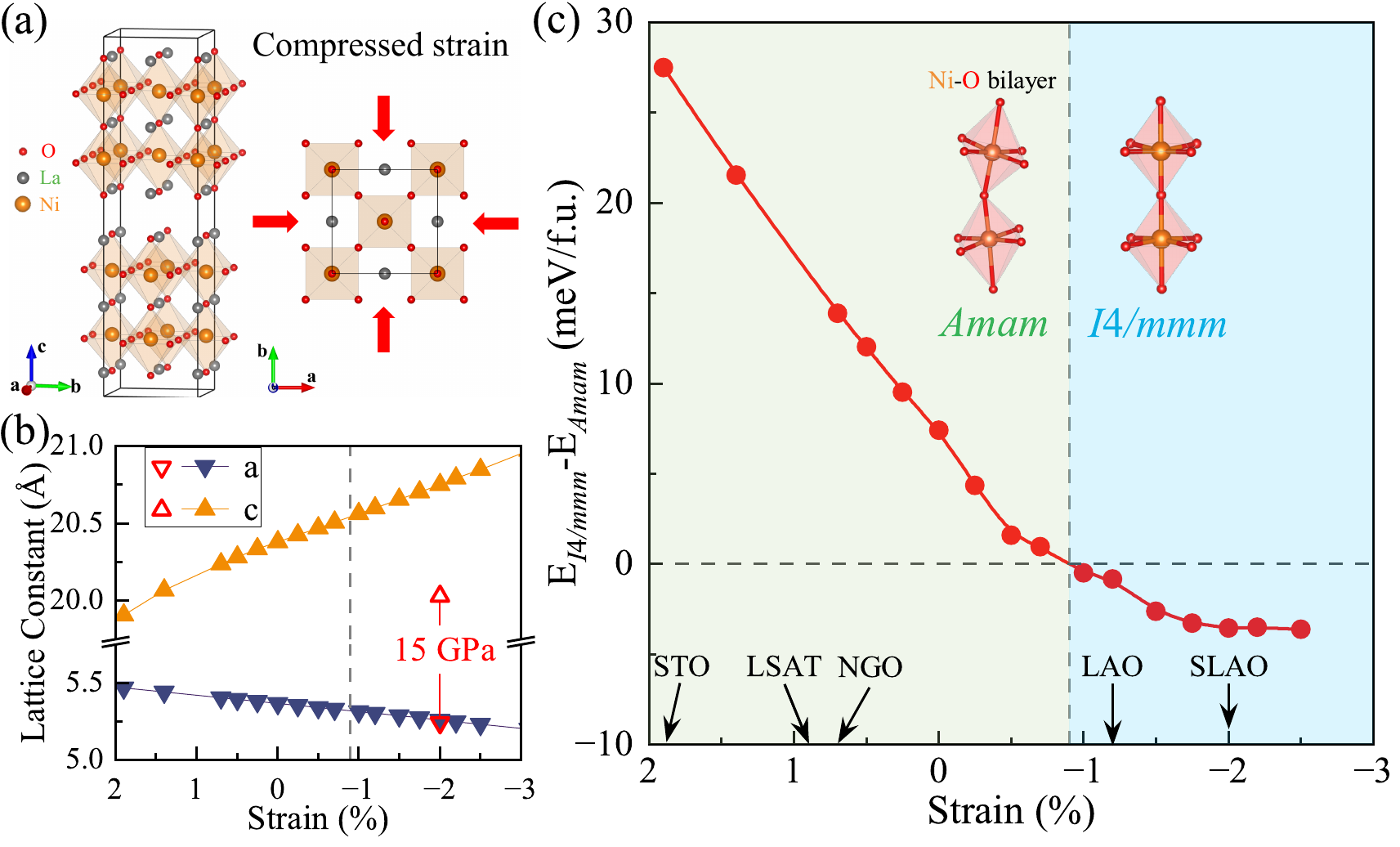}\\
	\caption{(a) Crystal structure of La\textsubscript{3}Ni\textsubscript{2}O\textsubscript{7} with in-plane compressive strain indicated by red arrows. (b) Evolution of in-plane and out-of-plane lattice parameters (a and c) under strain. (c) Energy difference E$_{I4/mmm}$-E$_{Amam}$ versus strain. SLAO, LAO, NGO, LSAT and STO denote commonly used substrates for growing La$_3$Ni$_2$O$_{7}$: SrLaAlO$_4$(001), LaAlO$_3$(001), NdGaO$_3$(001), (LaAlO$_3$)$_{0.3}$(Sr$_2$TaAlO$_6$)$_{0.7}$(001) and SrTiO$_3$(001), respectively \cite{RN1943,RN2003, RN1944, 2501.08204, 2501.06875, 2501.08022}. 
	}\label{energy}
\end{figure}

$\textit{Structural transition.}$---Fig.~\ref{energy}(a) illustrates the crystal structure of La$_{3}$Ni$_{2}$O$_{7}$, where biaxial compressive strain from substrates modifies the in-plane lattice parameters. Unlike previous approaches \cite{RN1703}, we simulate experimental strain conditions by fixing in-plane lattice constants while allowing c-axis relaxation during optimization. Due to the Poisson effect, in-plane compression leads to out-of-plane expansion (Fig.~\ref{energy}(b)). The primary structural difference between low-symmetry $Amam$ and tetragonal $I4/mmm$ phases lies in the interlayer Ni-O-Ni bond angles (inset of Fig.~\ref{energy}(c)). Strain-dependent energy calculations reveal that tensile strain stabilizes the $Amam$ phase, while compressive strain reduces the energy difference and activates a transition to $I4/mmm$ phase at $-0.9\%$ strain, consistent with recent scanning transmission electron microscopy (STEM) observations: $I4/mmm$ phase is stabilized on SrLaAlO$_4$ (SLAO, -2.0\%) and LaAlO$_3$ (LAO, -1.2\%) substrates, whereas the $Amam$ phase persists on NdGaO$_3$ (NGO, +0.7\%) and SrTiO$_3$ (STO, +1.9\%) substrates \cite{2501.06875, 2501.08204}. These findings contradict earlier predictions requiring over $-4.5\%$ strain to attain the high-symmetry phase \cite{RN1703}. Notably, hydrostatic pressure of 15 GPa triggering the structural transition induces $\sim-2\%$ in-plane compression (Fig.~\ref{energy}(b)), demonstrating epitaxial strain tunes the crystal structure more effectively than pressure.

\begin{figure}[!t]
	\centering
	\includegraphics[scale=0.58,angle=0]{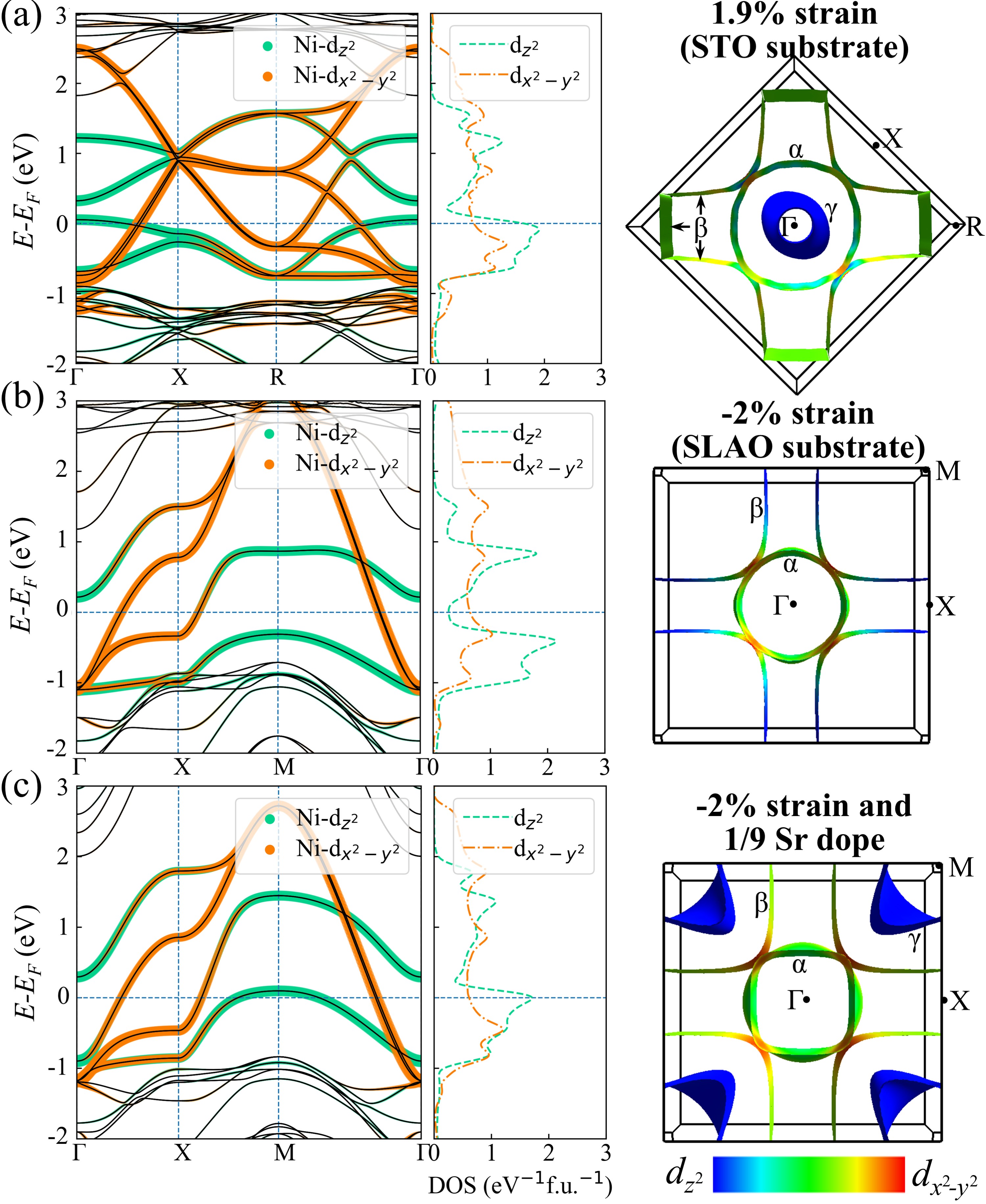}\\
	\caption{Projected band structures, density of states (DOS) and Fermi surfaces of La$_3$Ni$_2$O$_7$ under (a) $1.9\%$ tensile strain on STO substrate, (b) $-2\%$ compressive strain on the SLAO substrate and (c) $-2\%$ compressive strain with 1/9 Sr doping.}\label{band}
\end{figure}

$\textit{Band structures.}$---Fig.~\ref{band}(a) and (b) show the projected band structures and Fermi surface of La$_3$Ni$_2$O$_7$ under $1.9\%$ tensile and $-2\%$ compressive strain, respectively. Compared with the unstrained case, tensile strain raises the $d_{z^2}$ orbital energy levels, forming a new hole-like Fermi surface pocket ($\gamma$ pocket) around the $\Gamma$ point, resembling those calculated for high-symmetry bulk La$_3$Ni$_2$O$_{7}$ under high pressure \cite{RN1473}. Conversely, compressive strain lowers the $d_{z^2}$ levels, and the Fermi surfaces at $-2\%$ strain retains two primary pockets ($\alpha$ and $\beta$ pockets), similar to the unstrained configuration. These results agree with recent ARPES experiments \cite{2504.16372}. Notably, structural differences between $Amam$ and $I4/mmm$ phases lead to folding effects in bands and Fermi surfaces shown in Fig.~\ref{band}(a) compared to Fig.~\ref{band}(b). Additional results under various strain conditions are provided in Figs. S1 and S2 in the supplemental materials (SM) \cite{supp}.

Complementary STEM and energy-dispersive X-ray spectroscopy (EDS) measurements confirmed significant interfacial Sr diffusion with a length scale of $\sim1$ unit cell \cite{2501.09255}. To investigate this doping effect, we employed the virtual crystal approximation to simulate 1/9 Sr substitution for La. The resulting band structure (Fig.~\ref{band}(c)) shows an upward $\sim0.5$ eV in the bonding $d_{z^2}$ energy level, which across E$_F$ and reproduces the experimentally observed $\gamma$ pocket around M by ARPES measurements \cite{2501.09255}. These results demonstrate that interfacial Sr-induced hole doping plays a crucial role in reshaping the band structure of superconducting La$_3$Ni$_2$O$_{7}$ films.

\begin{table}[!t]
	\renewcommand\arraystretch{1.1}
	\caption{The tight-binding parameters for the two-orbital model of La$_3$Ni$_2$O$_{7}$ under strain or pressure (P), including $t_\parallel^x$ (in-plane nearest-neighbor (NN) $d_{x^2-y^2}$ hopping), $t_\parallel^z$ (in-plane NN $d_{z^2}$ hopping), $t_\parallel^{xz}$ (hybridization between in-plane $d_{x^2-y^2}$ and $d_{z^2}$ orbitals), $t_{\perp}^z$ (out-of-plane $d_{z^2}$ NN orbital hopping), $\epsilon_x$-$\epsilon_z$ (on-site energy differences of $d_{x^2-y^2}$ and $d_{z^2}$ orbitals) (all units in meV).}
	{\centering
		\begin{tabular}{lp{1cm}<{\centering}p{1.8cm}<{\centering}p{0.8cm}<{\centering}p{0.8cm}<{\centering}p{0.8cm}<{\centering}p{0.8cm}<{\centering}p{0.92cm}<{\centering}p{1cm}}
			\hline
			\hline
			& & Space group & t$_\parallel^x$               & t$_\parallel^z$               & t$_\parallel^{xz}$              & t$_{\perp}^z$ & $\epsilon_x-\epsilon_z$         \\
			\hline
			&Strain=$0\%$  & $Amam$ &  -397 & -73 &  189 & -618         & 765 \\
			&\multicolumn{1}{r}{$-1.2\%$}& $I4/mmm$                      & -461                        & -85                        & 206                        & -643             & 956 \\
			&\multicolumn{1}{r}{$-2\%$} & $I4/mmm$                      & -468                       & -78                        & 201                        & -650             & 1138 \\
			&\multicolumn{1}{r}{$-4.1\%$}& $I4/mmm$                      & -470                         & -63                       & 187                       & -667         & 1490  \\
			\hline
			&\multicolumn{1}{r}{P=10 GPa}& $Amam$                   & -475                        & -103                        & 230                         & -664         & 742 \\
			&\multicolumn{1}{r}{30 GPa}& $I4/mmm$                 & -522                        & -112                        & 265                       & -714        & 762 \\
			&\multicolumn{1}{r}{50 GPa}& $I4/mmm$                 & -556                        & -125                        & 286                        & -750          & 772 \\
			&\multicolumn{1}{r}{70 GPa}& $I4/mmm$                 & -582                        & -137                        & 303                        & -780          & 767 \\
			&\multicolumn{1}{r}{100 GPa}& $I4/mmm$                & -614                        & -152                        & 324                       & -819         & 750  \\
			&\multicolumn{1}{r}{150 GPa} & $I4/mmm$                & -645                        & -167                        & 351                        & -861         & 742\\
			\hline
			\hline	
	\end{tabular}}\label{table1}
\end{table}

Based on band structures, we get the key tight-binding parameters of the two-orbital model (Ni-$d_{x^2-y^2}$ and $d_{z^2}$ orbitals) under strain or pressure, derived via Wannier downfolding of DFT band structures in Table~\ref{table1}. While all parameters increase under pressure, compressive strain yields more intricate results. At $-2\%$ strain, the in-plane nearest-neighbor (NN) $d_{x^2-y^2}$ hopping $t_{\parallel}^x$ increases by $\sim16\%$, primarily attributed to the in-plane Ni-Ni bond distance contraction. In contrast, the out-of-plane $d_{z^2}$ orbital hopping $t_{\perp}^z$ shows minimal enhancement, 
as the interlayer Ni-Ni distance expands only slightly from 3.89 $\text{\AA}$ (unstrained) to 3.93 $\text{\AA}$ ($-2\%$ strain) and its enhancement is mainly due to the straightening of interlayer Ni-O-Ni bond angles to $180^\circ$, driven by the structural transition. The on-site energy difference between $d_{x^2-y^2}$ and $d_{z^2}$ orbitals ($\epsilon_x - \epsilon_z$) significantly increases under compressive strain, consistent with the downward shift of $d_{z^2}$ bands seen in Figs.~\ref{band} and S2. Furthermore, both the hybridization between in-plane $d_{x^2-y^2}$ and $d_{z^2}$ orbitals $t_\parallel^{xz}$ and in-plane NN $d_{z^2}$ hopping $t_\parallel^z$ initially increase, then gradually decrease with increasing compressive strain. This non-monotonic trend arises from competing effects: enhanced orbital overlap due to in-plane Ni-Ni bond contraction versus increased localization of $d_{z^2}$ bonding states as they move away from the E$_F$.

\begin{figure}[!t]
	\centering
	\includegraphics[scale=0.172,angle=0]{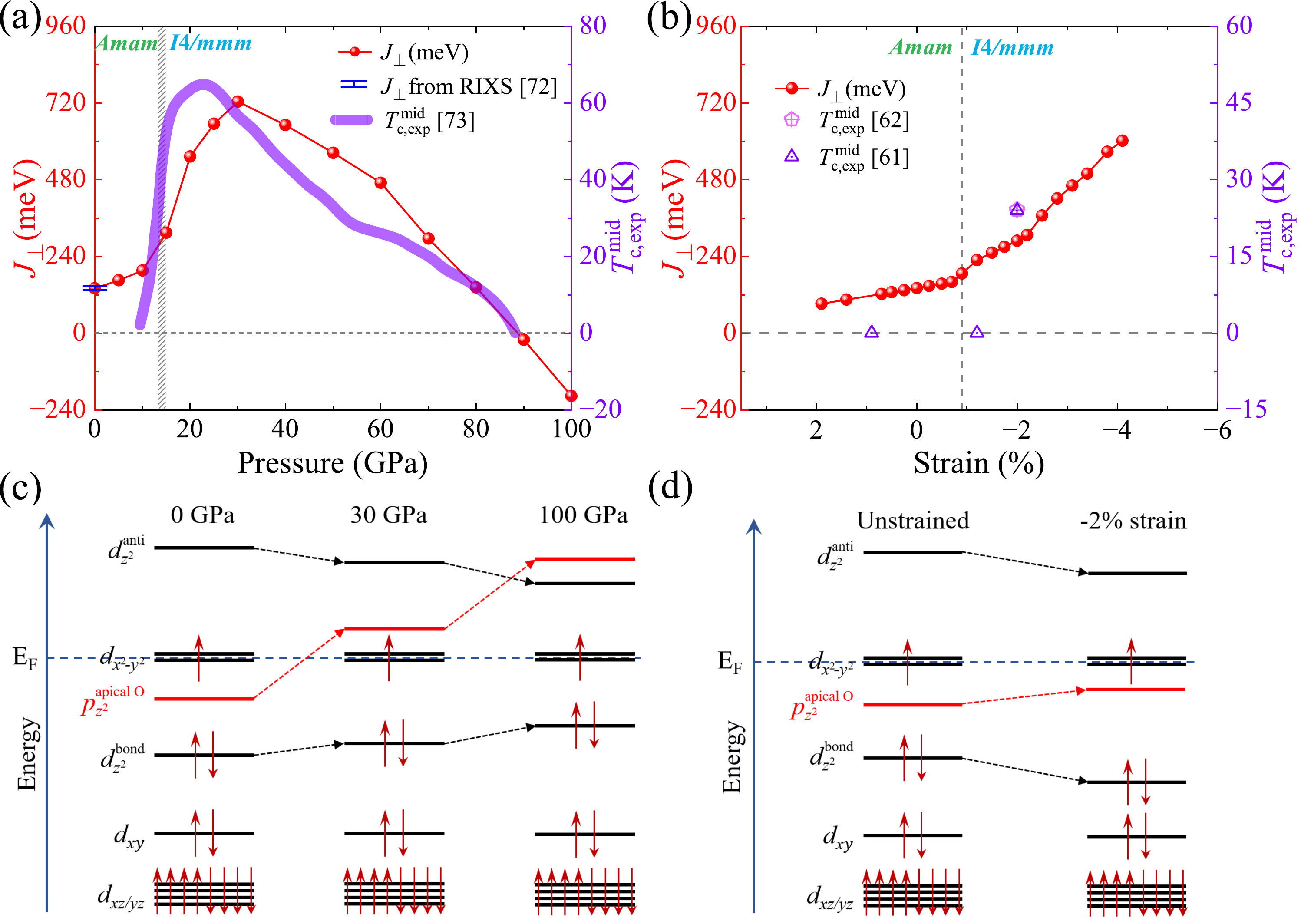}\\
	\caption{(a) Pressure- and (b) strain-dependent evolution of $J_{\perp}$ (red curves) and experimental superconducting $T_{\text{c,exp}}^{\text{ mid}}$ (purple curves and dots), with data from Refs.\cite{2404.11369,RN1555,RN1943,RN2003}. Orbital energies configurations of Ni-$d$ orbitals and apical O-$p_z$ orbitals under (c) 30 and 100 GPa and (d) 0 and $-2\%$ strain. Red arrows denote the distribution of 15 d-electrons in two Ni\textsuperscript{2.5+} ions. Blue dashed lines denote the Fermi level (E$_F$) . Bonding ($d_{z^2}^{\mathrm{bond}}$) and antibonding ($d_{z^2}^{\mathrm{anti}}$) states emerge through adjacent Ni-$d_{z^2}$ orbitals.}\label{jz}
\end{figure}

$\textit{Interlayer antiferromagnetic coupling.}$---Largest hopping parameter t$_{\perp}^z$ induces strong interlayer antiferromagnetic coupling $J_{\perp}$, which is crucial in governing properties of La$_3$Ni$_2$O$_{7}$. $J_{\perp}$ under pressure or strain is shown in Fig.~\ref{jz}. Under pressure up to 100 GPa, $J_{\perp}$ ($>0$ means antiferromagnetic coupling) first increases with a slow growth rate below 15 GPa, followed by a sharp rise between 15-30 GPa, peaking at $\sim$30 GPa, attributed to the strengthened $d_{z^2}$ orbital overlap due to 180° Ni-O-Ni bond alignment in the $I4/mmm$ phase. Beyond 30 GPa, $J_{\perp}$ gradually decreases and becomes negative around 90 GPa, signaling a transition from antiferromagnetism to ferromagnetism. The $J_{\perp}$ estimated by resonant inelastic X-ray scattering (RIXS) experiments at ambient pressure aligns well with calculations (Fig.~\ref{jz}(a)) \cite{RN1555}.

Under epitaxial strain, $J_{\perp}$ varies monotonically within $-4.1\%$ to $2\%$ strain range: increasing under compressive strain and decreasing under tensile strain. Below $-0.9\%$, $J_{\perp}$ of the $I4/mmm$ phase shows a steeper rise, consistent with RIXS observations reporting a $\sim20\%$ enhancement at $\sim -2\%$ strain compared to the unstrained case \cite{2502.03178}, further validating our results.

Experimental mid-transition superconducting temperatures $T_{\text{c,exp}}^{\text{ mid}}$, defined as the temperature where electrical resistance drops to $50\%$ of its initial transition value, are also plotted in Figs.~\ref{jz}(a) and (b). Under pressure, $T_{\text{c,exp}}^{\text{ mid}}$ displays a dome-shaped evolution peaking near 25 GPa \cite{2404.11369}, closely mirroring the pressure dependence of $J_{\perp}$ in Fig.~\ref{jz}(a). In strained systems, superconductivity is only observed at $-2\%$ strain using SLAO substrates \cite{RN1943,RN2003,2501.08022}, where $J_{\perp}$ is comparable to that at 15 GPa. This suggests that large compressive strain may be required to achieve sufficient $J_{\perp}$ for superconductivity.

To investigate the pressure- and strain-dependence of $J_{\perp}$, we systematically analyze the governing parameters. The Ni-Ni interlayer superexchange primarily arises from $d_{z^2}$ orbital interactions mediated by apical oxygen atoms, expressed as (see SM for more details \cite{supp}):

\begin{equation}
	J_{\perp} \propto |V_{pd}|^4 \frac{E_{d_{z^2}}^{\text{bond}} -E_{d_{z^2}}^{\text{anti}}}{(E_{p_z}^{\text{apical O}} - E_{d_{z^2}}^{\text{bond}})(E_{d_{z^2}}^{\text{anti}} - E_{p_z}^{\text{apical O}}) }, \label{eq1}
\end{equation} 
where $V_{pd}$ is the hopping integral between apical O $p_z$ and Ni $d_{z^2}$ orbitals, which is obtained via Wannier function fitting. $E_{d_{z^2}}^{\text{anti}}$, $E_{d_{z^2}}^{\text{bond}}$ and $E_{p_z}^{\text{apical O}}$ represent average energy levels of antibonding and bonding states of Ni-$d_{z^2}$ orbitals, and apical O-$p_{z}$ orbitals, respectively. Calculated orbital energy levels and $V_{pd}$ can be found in Table S2 \cite{supp}, and orbital energies configurations are shown in Figs.~\ref{jz}(c) and (d). 

Under increasing pressure, the $E_{d_{z^2}}^{\text{bond}}$ gradually increases while $E_{d_{z^2}}^{\text{anti}}$ decreases, reflecting progressive metallization of the $d_{z^2}$ bands \cite{RN1300}. Simultaneously, the upward shift of $E_{p_z}^{\text{apical O}}$ induces a substantial enhancement in the second term of Eq.~\ref{eq1} from 0 to 30 GPa, as detailed in Table S2 \cite{supp}. These combined effects drive the continuous increase in $J_{\perp}$ within the 0-30 GPa range. The rising $E_{p_z}^{\text{apical O}}$ eventually surpasses $E_{d_{z^2}}^{\text{anti}}$ at $\sim90$ GPa, resulting in a reduction and subsequent sign reversal of the second term in Eq. \ref{eq1}, thereby triggering a transition from interlayer antiferromagnetism to ferromagnetism. 

Under compressive strain, both $E_{d_{z^2}}^{\text{bond}}$ and $E_{d_{z^2}}^{\text{anti}}$ decrease concurrently, consistent with the downward shift of the $d_{z^2}$ bands shown in Fig.~\ref{band}. Although $E_{p_z}^{\text{apical O}}$ exhibits a minor increase, it remains positioned between the bonding and antibonding $d_{z^2}$ levels. The calculated results demonstrate that both terms in Eq. \ref{eq1} increase under compressive strain (Table S1), collectively accounting for the monotonic enhancement of $J_{\perp}$ observed in Fig.~\ref{jz}(b).

\begin{figure}[!t]
	\centering
	\includegraphics[scale=0.29,angle=0]{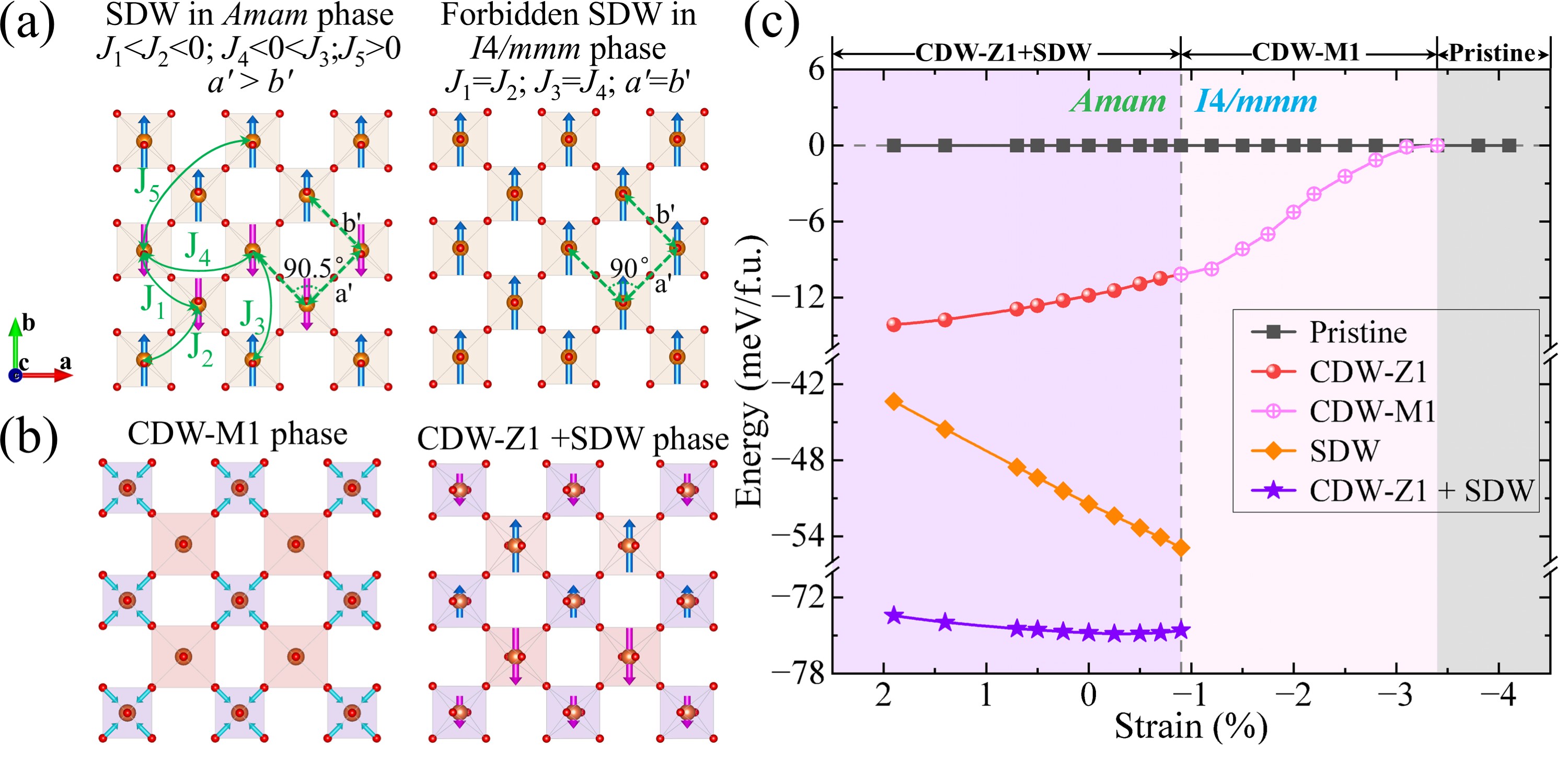}\\
	\caption{(a) In-plane magnetic structures of the SDW state in the $Amam$ phase and the AFM-A state in the $I4/mmm$ phase. Blue and pink arrows indicate spin orientations. Green arrows denote Ni-Ni magnetic couplings ($J_1$-$J_5$). $a^{\prime}$ and $b^{\prime}$ are in-plane distance of adjacent Ni atoms. (b) In-plane schematics of CDW-M1 and coexisting CDW-Z1 + SDW phases. Cyan arrows denote oxygen displacements; expanded and contracted octahedra are colored light red and purple, respectively. (c) Energy evolution of CDW, SDW, and CDW + SDW phases under strain relative to the pristine phase. }\label{cdw}
\end{figure}

$\textit{Density wave orders.}$---The pressure-induced phase diagram of La$_3$Ni$_2$O$_{7}$ reveals the emergence of both CDW and stripe-type SDW orders \cite{RN1346,RN1571,RN1503,RN1384,RN1417, RN1555,2401.12635, 2402.03952,RN1554,2410.15298,2503.05287,2503.09288}. Strain-dependent behaviors of CDW and SDW are systematically investigated in this section. Fig.~\ref{cdw}(a) presents two representative CDW structures (CDW-Z1 and CDW-M1) previously proposed, which originate from the $Amam$ and $I4/mmm$ pristine phases, respectively \cite{RN1596}. Both phases feature alternating in-plane expansion and contraction of oxygen octahedra within each layer (illustrated by light red and purple octahedra in Fig.~\ref{cdw}(b)). The energy of CDW phase in Fig.~\ref{cdw}(c) decreases under tensile strain but increases under compressive strain, with structural distortion completely suppressed at $\sim -3.3\%$ strain, similar to experimental observations of pressure-induced CDW suppression \cite{RN1356,RN1417,RN1300}.

For SDW, we focus on the double-stripe SDW configuration in the $Amam$ phase (Fig.~\ref{cdw}(a)). Its unequal in-plane NN Ni-Ni distances ($a^{\prime} > b^{\prime}$) yields a distorted non-square Ni lattice and a hierarchy of magnetic coupling: $J_1 < J_2 < 0$, $J_4 < 0 < J_3$, and $J_5 > 0$, collectively stabilizing the stripe SDW order (Table S3 of SM \cite{supp}). As shown in Fig.~\ref{cdw}(c), compressive strain thermodynamically stabilizes SDW while tensile strain suppresses it, opposite to the CDW trend. In contrast, the $I4/mmm$ phase maintains $a^{\prime} = b^{\prime}$, producing a square Ni lattice with $J_1 = J_2$ and $J_3 = J_4$, which forbids SDW and favors an AFM-A ground state with interlayer antiferromagnetism and ferromagnetism. This explains the absence of SDW in superconducting films with compressive strain beyond $-0.9\%$ \cite{RN1943,RN2003}.

Coexistence of CDW and SDW, reported in systems such as La$_4$Ni$_3$O$_8$ \cite{RN1396,RN1564}, doped La$_2$NiO$_4$ \cite{RN1565,RN1566,RN1567}, La$_4$Ni$_3$O$_{10}$ \cite{RN1563}, and Cr single crystals \cite{RN1533}, is also energetically favored in La$_3$Ni$_2$O$_{7}$ according to Fig.~\ref{cdw}(c), but only persists above $-0.9\%$ strain where SDW survives. As shown in Fig. S4 \cite{supp}, oxygen octahedral distortions split magnetic moments on central Ni atoms, generating a spin-charge-orbital ordering \cite{RN1596,RN1816}. The resulting magnetic pattern of CDW-Z1 + SDW phase in Fig.~\ref{cdw}(b) aligns with magnetic moment stripes observed via ambient-pressure magnetic neutron scattering \cite{2503.05287}.

$\textit{Discussion.}$---Compring pressure- and strain-driven superconductivity offers insights into key governing factors. A central yet unresolved issue is the role of the $\gamma$ pocket originating from $d_{z^2}$ bonding states. While this pocket emerges under pressure coinciding with superconductivity and is theoretically associated with s-wave pairing \cite{RN1331,RN2004}, it is absent under compressive strain, where $d_{z^2}$ states shift downward as evidenced by calculations in Fig.~\ref{band}(b) and ARPES measurements \cite{2504.16372}. Conversely, tensile strain creates $\gamma$ pocket without inducing superconductivity \cite{RN1943}, suggesting that metallization of $d_{z^2}$ bonding states alone is insufficient for superconductivity. The situation is more intricate by Sr diffusion near interfaces. 1/9 Sr hole doping can locally lower the E$_F$, reintroducing $d_{z^2}$ pockets (Fig.~\ref{band}(c)). However, this effect is spatially confined: only interfacial unit cells near substrates maintain metallic, whereas distant regions become insulating \cite{2501.09255}. These findings highlight the need for further experiments, such as employing different substrates to eliminate interface doping, or applying larger compressive strain to further lower $d_{z^2}$ energy levels.

The structural transition to $I4/mmm$ phase appears nonessential for superconductivity except for its suppression of SDW (Fig.~\ref{cdw}). Notably, the structural transition and superconducting onset are not synchronized. Our calculations indicate a structural transition at $\sim-0.9\%$ strain, and experiments show that $-1.2\%$ strain with $I4/mmm$ phase fails to induce superconductivity \cite{2501.08204}. Furthermore, superconductivity is absent in various ambient-pressure systems with 180° out-of-plane Ni-O-Ni bond angles under ambient pressure, including 2222-La$_3$Ni$_2$O$_{7}$ \cite{2501.14202}, 1313-La$_3$Ni$_2$O$_{7}$ \cite{RN1516}, 1212-La$_5$Ni$_3$O$_{11}$ \cite{2502.01018}, and the tetragonal La$_4$Ni$_3$O$_{10}$ \cite{2501.12647}.

Tight-binding analysis (Table~\ref{table1}) further reveals that $t_{\perp}^z$ shows no significant enhancement in the 10-20 GPa or -2\% strain regimes where superconductivity emerges. Although $t_{\perp}^z$ increases beyond 30 GPa, superconductivity concurrently weakens \cite{2501.12647}, suggesting $t_{\perp}^z$ has minimal impact on superconductivity or its magnitude under ambient pressure is already sufficient, with other factors being responsible for suppressing superconductivity. Combined with our previous calculations based on density matrix renormalization group and thermal tensor networks, $t_\parallel^{xz}$, $t_\parallel^{x}$ and $\epsilon_x$-$\epsilon_z$ are critical parameters \cite{2311.12769}. Among these, $t_\parallel^{xz}$ remains within 0.19-0.35 eV under both pressure and strain, which acts synergistically with large $J_{\perp}$ to stabilize a Hund's coupling-driven superconductivity regime. The large $t_\parallel^{x}$ favors strong pairing with $T_c \approx 0.03t_\parallel^{x}$. In addition, $\epsilon_x$-$\epsilon_z$ below 0.8 eV dramatically enhances superconductivity, while further increase exhibits negligible effects.

The interlayer antiferromagnetic coupling is widely regarded as a key driver of superconducting pairing of La$_3$Ni$_2$O$_{7}$, though the precise mechanism in La$_3$Ni$_2$O$_{7}$ remains debated \cite{2306.07275, RN1333, RN1355, RN1334, RN1399,RN2083,RN1453}. Our DFT calculations (Fig.~\ref{jz}) corroborate this picture, showing enhanced superconductivity under pressure or strain in parallel with increasing $J_{\perp}$, highlighting a strong correlation between interlayer coupling and superconductivity. Moreover, functional renormalization group analyses indicate that increasing pressure gradually suppresses spin fluctuations, consistent with suppressed $J_{\perp}$ \cite{RN2004}.

Superconductivity in bulk La$_3$Ni$_2$O$_{7}$ under pressure is observed following the suppression of competing density wave orders \cite{RN1571,RN1384,RN1417}, while non-superconducting films retain collinear spin order and charge anisotropy \cite{RN1849}. Our strain-dependent calculations in Fig.~\ref{cdw} demonstrate that superconductivity coincides with the suppression of both SDW and CDW at -2\% strain, reinforcing the picture of competing orders. To comprehensively map the evolution of these orders under strain, future studies using techniques such as transport measurements, neutron scattering studies, muon spin relaxation, and STEM are highly desirable.

Figs.~\ref{jz} and \ref{cdw} collectively suggest that enhanced compressive strain could improve $T_c$ by simultaneously boosting $J_{\perp}$ and suppressing CDW. We propose possible high-symmetry cubic or tetragonal substrates, such as NdAlO$_3$ ($-2.5\%$ strain) \cite{geller1956crystallographic}, SrGd$_2$Al$_2$O$_7$ ($-2.5\%$) \cite{fava1975phases}, and AlAsO$_4$ ($-6.2\%$) \cite{strada1934struttura}, as promising to enhance superconductivity. These substrates can provide substantial biaxial compressive strain while maintaining the $I4/mmm$ symmetry in films. In contrast, low-symmetry substrates may strengthen competing orders. For instance, La$_3$Ni$_2$O$_{7}$ films grown on orthorhombic YAlO$_3$ ($a \neq b$), despite experiencing $-4.1\%$ compressive strain, show no superconductivity \cite{RN1493}, likely due to structural distortion reinforcing SDW/CDW.


$\textit{Conclusion.}$---In summary, systematic DFT calculations reveal the key structural, electronic and magnetic factors unifying strain- and pressure-driven superconductivity in La$_3$Ni$_2$O$_{7}$. The $I4/mmm$ phase transition at $-0.9\%$ in-plane compressive strain precedes superconductivity at $-2\%$ strain, accompanied by a downward shift of Ni-$d_{z^2}$ orbitals. Interfacial Sr diffusion reconstructs $d_{z^2}$ Fermi pockets. Tight-binding analysis highlights the importance of $t_\parallel^{xz}$, $t_\parallel^{x}$ and $\epsilon_x$-$\epsilon_z$ in mediating pairing. Magnetic calculations demonstrate that change in $J_\perp$, driven by apical O-$p_z$ orbital energy shifts, closely tracks the evolution of $T_c$ under pressure and strain. Notably, $-2\%$ strain and 15 GPa hydrostatic pressure achieve comparable $J_\perp$, $T_c$ and in-plane lattice constants, establishing a direct strain-pressure correspondence. Furthermore, compressive strain suppresses both CDW and SDW orders. These results suggest that superconductivity emerges from joint effect of density waves suppression and enhanced $J_\perp$. Our calculations are in quantitative agreement with several experimental observations. This work offers guidance for strain engineering in nickelate superconductors, advancing both mechanistic understanding and materials optimization.

\textit{Note added}.--- During the preparation of this work, we became aware of several independent studies investigating band structures of strained La$_3$Ni$_2$O$_{7}$ based on DFT \cite{RN1948,2501.14665,2502.04255,2503.10902} or employing its tight-binding models \cite{2502.01624,2503.17223}.

		\textit{Acknowledgments}.--- 
		This work is supported in part by the Strategic Priority Research Program of the Chinese Academy of Sciences (Grant No. XDB28000000), the National Natural Science Foundation of China (Grant No.11834014), and the Innovation Program for Quantum Science and Technology (Grant No. 2021ZD0301800). B.G. is supported in part by the National Natural Science Foundation of China (Grant No. 12074378), the Chinese Academy of Sciences (Grants No. YSBR-030, No. JZHKYPT-2021-08, No. XDB33000000), Beijing Municipal Science and Technology Commission (Grant No. Z191100007219013).


%

\newpage
\clearpage
\onecolumngrid
\mbox{}
\begin{center}
	{\large Supplementary Materials for} $\,$ \\	
	\bigskip	
	\textbf{\large{Unifying Strain-driven and Pressure-driven Superconductivity in La$ _{3} $Ni$ _{2} $O$ _{7} $: Suppressed charge/spin density waves and enhanced interlayer coupling}} \\		
	Yi \textit{et al}.
\end{center}
\date{\today}
\setcounter{section}{0}
\setcounter{table}{0}
\setcounter{figure}{0}
\setcounter{equation}{0}
\renewcommand{\thetable}{S\arabic{table}}
\renewcommand{\theequation}{S\arabic{equation}}
\renewcommand{\thefigure}{S\arabic{figure}}
\renewcommand{\thesection}{S\arabic{section}}
\setcounter{secnumdepth}{3}

\section{Methods}
The Vienna Ab initio Simulation Package (VASP) was employed for density functional theory (DFT) calculations \cite{RN142}. The plane-wave cutoff energy was set to 520 eV, with atomic force and total energy are set to 1 meV/\AA\ and $10^{-7}$ eV/atom, respectively. A $\Gamma$-centered 12×12×12 Monkhorst-Pack k-point grid was used in reciprocal space for self-consistent calculations of the primitive cell. The exchange-correlation energy was treated using the Perdew-Burke-Ernzerhof (PBE) scheme within the generalized gradient approximation (GGA) \cite{RN143}. We employed virtual crystal approximation (VCA) to model 1/9 Sr substitution for La \cite{RN2090}. Recent theoretical studies indicated that Hubbard U values should not exceed 2 eV to better capture the experimentally observed spin density wave orders \cite{RN1571,RN1503,RN1555,2402.03952,2410.15298,2401.12635,2503.05287,2503.09288,RN1817,RN1816,RN1814}. So Most calculations in the main text and supplementary materials adopt U= 1.5-2 eV, except that U = 4 eV is used for Figs. S1 and S2, and Table I. It should be noted that different U do not affect corresponding conclusions as discussed in section S2. The Wannier90 package \cite{RN1656} was utilized to construct maximally localized Wannier functions and downfold band structures. Two-orbital models are established by including Ni $d_{x^2-y^2}$ and $d_{z^2}$ orbitals. Magnetic exchange couplings were calculated based on the magnetic force theorem using the TB2J package \cite{RN2089}.

Experiments have identified two high-symmetry phases of La$_3$Ni$_2$O$_{7}$ under high pressure—-the orthorhombic $Fmmm$ and tetragonal $I4/mmm$ phases \cite{RN1473,RN1300}, but their structural differences are minimal. First-principles DFT calculations indicate that the key distinction lies in the in-plane lattice constants, where the $Fmmm$ phase exhibits a 0.01 $\text{\AA}$ contraction along the a-axis relative to the b-axis ($a < b$). This induces negligible changes in both magnetic and electronic properties between them. Additionally, commonly used orthorhombic substrates for La$_3$Ni$_2$O$_{7}$ thin films, such as SrLaAlO$_4$ (SLAO), LaAlO$_3$ (LAO), NdGaO$_3$ (NGO), (LaAlO$_3$)$_{0.3}$(Sr$_2$TaAlO$_6$)$_{0.7}$ (LSTO) and SrTiO$_3$ (STO), impose the in-plane lattice constraint with $a = b$ \cite{RN1943,RN2003, RN1944, 2501.08204, 2501.06875, 2501.08022}. Therefore, we adopt $I4/mmm$ as the representative high-symmetry phase in our calculations.

\section{Electronic structures}

\begin{figure}[!h]
	\centering
	\includegraphics[scale=0.5,angle=0]{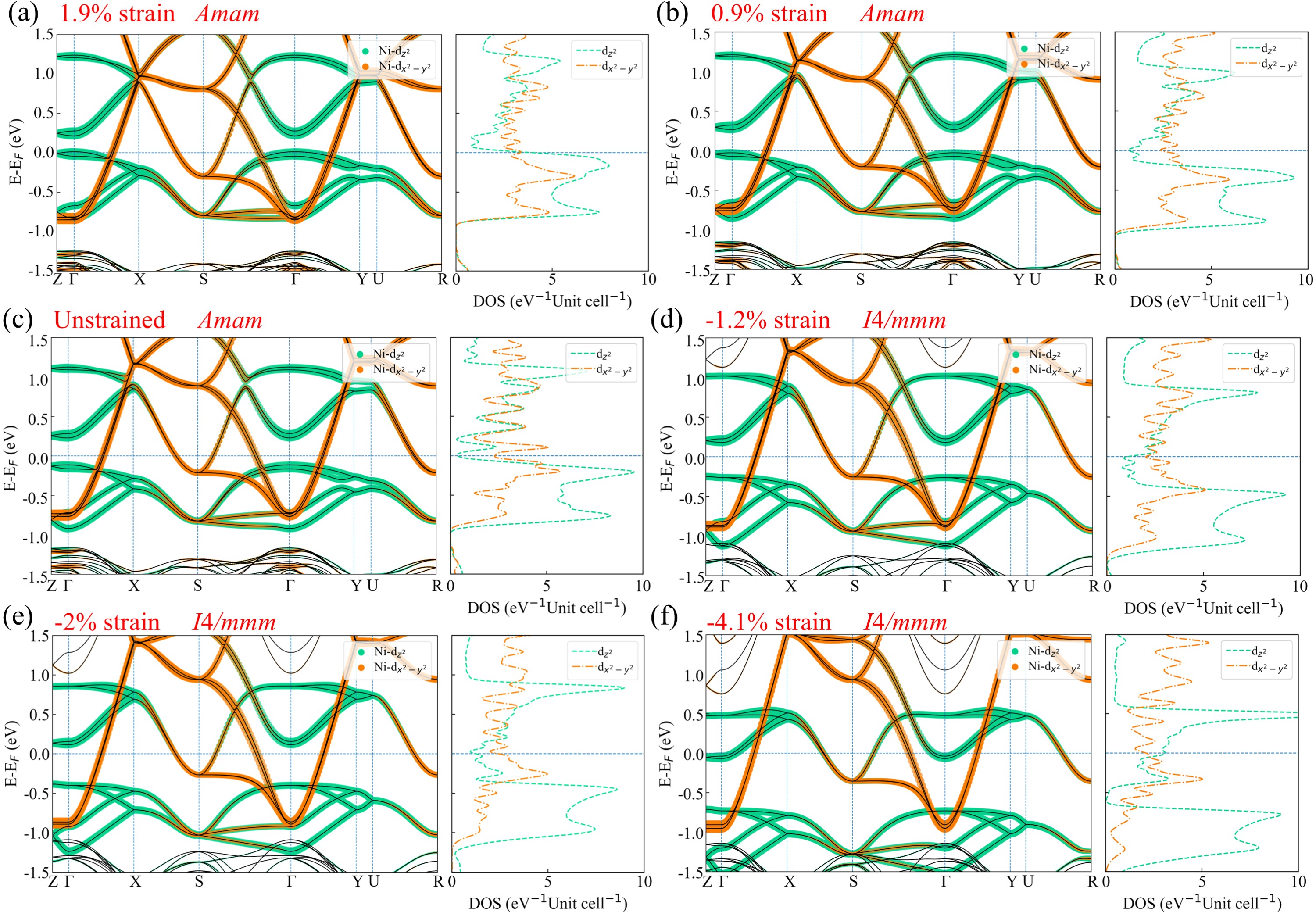}\\
	\caption{Band structures projected onto Ni-$d_{x^2-y^2}$ and $d_{z^2}$ orbitals for La\textsubscript{3}Ni\textsubscript{2}O\textsubscript{7} under (a) $1.9\%$ , (b) $0.9\%$, (c) $0\%$, (d) $-1.2\%$, (e) $-2\%$, and (f) $-4.1\%$ strain, respectively.}\label{band-dos-all-strain}
	\vspace{0cm}
\end{figure}

\begin{figure}[!h]
	\centering
	\includegraphics[scale=0.5,angle=0]{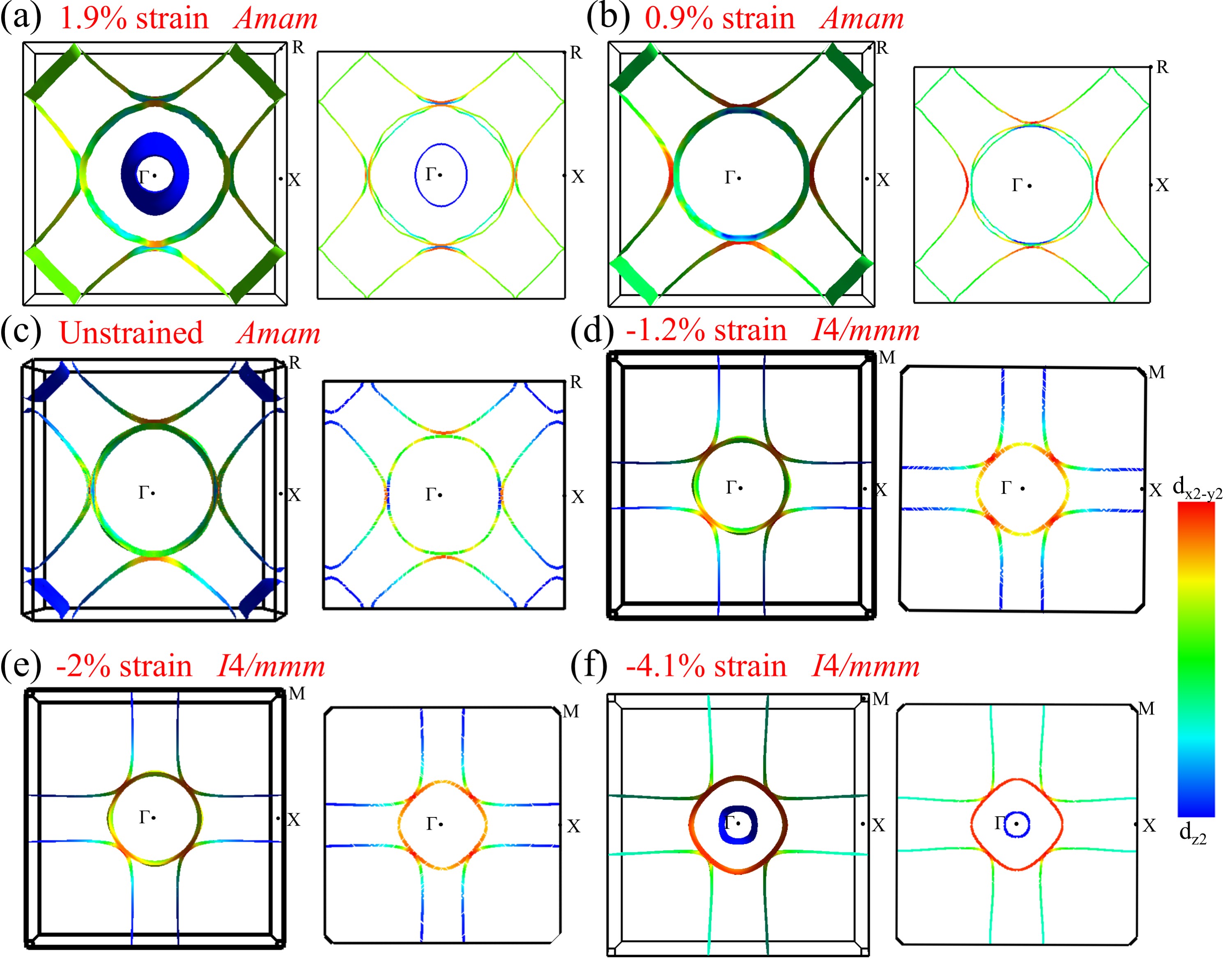}\\
	\caption{Fermi surfaces projected onto Ni-$d_{x^2-y^2}$ and $d_{z^2}$ orbitals for La\textsubscript{3}Ni\textsubscript{2}O\textsubscript{7} under (a) $1.9\%$, (b) $0.9\%$, (c) $0\%$, (d) $-1.2\%$, (e) $-2\%$, and (f) $-4.1\%$, respectively.}\label{FS-all-strain}
		\vspace{2cm}
\end{figure}

Band structures and Fermi surfaces projected onto Ni-$d_{x^2-y^2}$ and $d_{z^2}$ orbitals calculated with U = 4 eV for La\textsubscript{3}Ni\textsubscript{2}O\textsubscript{7} over a strain range of $1.9\%$ to $-4.1\%$ are shown in Figs.~\ref{band-dos-all-strain} and \ref{FS-all-strain}. Compared with the unstrained case, tensile strain elevates the $d_{z^2}$ orbital energy levels, introducing a new hole-like Fermi surface pocket around the $\Gamma$ point, resembling the high-pressure band features of high-symmetry phase. Conversely, compressive strain lowers the $d_{z^2}$ orbital energy levels. The Fermi surfaces at $-2\%$ strain remain similar to the unstrained case, featuring $\Gamma$- and M-centered pockets. At -4.1\% strain, bonding electronic states of $d_{z^2}$ orbitals approach the Fermi level, forming an additional $\Gamma$-centered pocket. Notably, these features are robust with respect to the choice of U = 2 or 4 eV as seen in Figs.~\ref{band-dos-all-strain} and \ref{FS-all-strain} and Fig. 2. Additionally, varying U values has minimal influence on all tight-binding hopping parameters within the two-orbital model, with only minor reductions in $\epsilon_x$-$\epsilon_z$ for lower U values (Table I of main text and Table~\ref{tableS1}).

\section{Density waves}

Strain-dependent Ni magnetic moments in different density wave orders are shown in Fig.~\ref{dw-magnetism}. In charge density wave (CDW) states, oxygen octahedra undergo in-plane expansion and contraction distortions, which are respectively colored light red and purple in Fig. 4(b). The structural distortion induced by CDW leads to two inequivalent Ni sites with split magnetic moments. This moment splitting disappears as CDW is suppressed, vanishing entirely at $-3.3\%$ strain.

\section{magnetic properties}

To investigate the pressure- and strain-dependent behavior of $J_{\perp}$, we systematically analyze the governing parameters. The Ni-Ni interlayer superexchange primarily arises from $d_{z^2}$ orbital interactions mediated by apical oxygen atoms, expressed as: 
$$J_{\perp} = \frac{1}{4A} \sum_{d_1, p, p', d_2} |V_{pd}|^2 J_{p'd_2},$$

where 
$$A = \left[ \frac{1}{(E_{d_1 d_1'}^{\uparrow \uparrow})^2} - \frac{1}{(E_{d_1 d_1'}^{\uparrow \downarrow})^2} \right]$$
is considered as constant under external perturbations \cite{RN2087}. Here, $E_{d_1 d_1'}^{\uparrow \uparrow}$ and $E_{d_1 d_1'}^{\uparrow \downarrow}$ represent the energies of two $d_{z^2}$ electrons (at sites $d_1$ and $d_2$) with parallel and antiparallel spins, respectively. $V_{pd}$ denotes the hopping integral between apical O $p_z$ and Ni $d_{z^2}$ orbitals. The $J_{p'd_2}$ term derived via the Schrieffer-Wolff transformation follows: 
$$J_{p'd_2} = 2|V_{pd}|^2 \left( \frac{1}{E_{p_z}^{\text{apical O}} - E_{d_{z^2}}^{\text{bond}}} + \frac{1}{E_{d_{z^2}}^{\text{anti}} - E_{p_z}^{\text{apical O}}} \right),$$ 
where $E_{d_{z^2}}^{\text{anti}}$, $E_{d_{z^2}}^{\text{bond}}$ and $E_{p_z}^{\text{apical O}}$ represent energy levels of antibonding and bonding states of Ni-$d_{z^2}$ orbitals, and apical O-$p_{z}$ orbitals, respectively.
This yields 

\begin{equation}
	J_{\perp} \propto |V_{pd}|^4 \frac{E_{d_{z^2}}^{\text{bond}} -E_{d_{z^2}}^{\text{anti}}}{(E_{p_z}^{\text{apical O}} - E_{d_{z^2}}^{\text{bond}})(E_{d_{z^2}}^{\text{anti}} - E_{p_z}^{\text{apical O}}) }.
\end{equation} 

Orbital energy levels are calculated by $E_i = \int g_i(E) E \, dE / \int g_i(E) \, dE$, where $g_i(E)$ represents the electronic density of states for orbital $i$. To determine the hopping integral $V_{pd}$ between apical O $p_z$ and Ni $d_{z^2}$ orbitals, maximally localized Wannier functions are constructed using all Ni-$d$ orbitals and O-$p$ orbitals. Calculated results can be found in Table~\ref{tableS2}. The calculated orbital energies are visualized as electronic energy configurations in Figs. 3(c) and (d) of the main text. 

Calculated Ni-Ni magnetic couplings ($J_1$-$J_5$, green arrows in Fig. 4(b)) for La$_3$Ni$_2$O$_{7}$ in the unstrained and -2\% strain are shown in Table~\ref{tableS3}. For the $Amam$ phase, its low symmetry results in unequal in-plane nearest-neighbor (NN) Ni-Ni distances (a$^{\prime} > b^{\prime}$) and a non-square Ni lattice arrangement. This yields a complex hierarchy of magnetic couplings: $J_1<J_2<0$, $J_4<0<J_3$, and $J_5>0$, collectively stabilizing the stripe SDW order shown in Fig. 4(a). In contrast, the high-symmetry $I4/mmm$ phase preserves equivalent in-plane Ni-Ni distances (a$^{\prime}$=b$^{\prime}$) and square Ni lattice. With coupling constants satisfying $J_1=J_2$ and $J_3=J_4$, the SDW order is forbidden, leaving an AFM-A state characterized by interlayer antiferromagnetism and intralayer ferromagnetism.

\begin{figure}[!h]
	\centering
	\includegraphics[scale=1,angle=0]{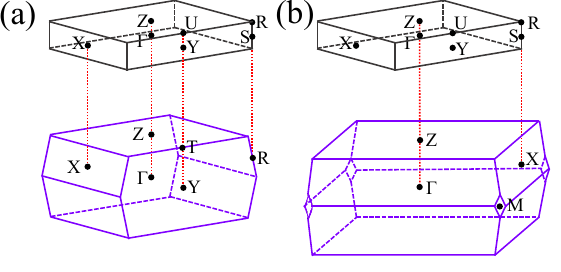}\\
	\caption{ Brillouin zone (BZ) of Bravais and primitive cells for (a) $Amam$ and (b) $I4/mmm$ phases. Bravais and primitive cells depicted by black and purple lines, respectively.}\label{BZ}
\end{figure}

\begin{table}[!h]
	\caption{Tight-binding parameters for the two-orbital model of La$_3$Ni$_2$O$_{7}$ under $-2\%$ strain for different Hubbard U, including $t_\parallel^x$ (in-plane NN $d_{x^2-y^2}$ orbital hopping), $t_\parallel^z$ (in-plane NN $d_{z^2}$ orbital hopping), $t_\parallel^{xz}$ (hybridization between in-plane $d_{x^2-y^2}$ and $d_{z^2}$ orbitals), $t_{\perp}^z$ (out-of-plane NN $d_{z^2}$ orbital hopping), $\epsilon_x$-$\epsilon_z$ (on-site energy differences of $d_{x^2-y^2}$ and $d_{z^2}$ orbitals).}
	{\centering
		\begin{tabular}{lp{1cm}<{\centering}p{1.2cm}<{\centering}p{1.2cm}<{\centering}p{1.2cm}<{\centering}p{1.2cm}<{\centering}p{1.2cm}<{\centering}p{1.8cm}<{\centering}p{1.5cm}}
			\hline
			\hline
			& Strain & U (eV) & t$_\parallel^x$ (eV)               & t$_\parallel^z$ (eV)           & t$_\parallel^{xz}$ (eV)        & t$_{\perp}^z$ (eV) & $\epsilon_x-\epsilon_z$ (eV)         \\
			\hline

			&-2\% & 4                      & -0.468                       & -0.078                        & 0.201                        & -0.65              & 1.138 \\
			&-2\% & 2                      & -0.474                        & -0.075                        & 0.196                        & -0.633             & 0.947 \\
			\hline
			\hline	
	\end{tabular}}\label{tableS1}
\vspace{0cm}
\end{table}

\begin{table}[!h]
	\caption{Calculated orbital energies of Ni and apical O atoms and hopping integral $V_{pd}$ between apical O $p_z$ and Ni $d_{z^2}$ orbitals under strain or pressure.}
	{\centering
		\begin{tabular}{lp{2cm}<{\centering}p{2cm}<{\centering}p{2cm}<{\centering}p{2.3cm}<{\centering}p{5cm}<{\centering}p{2cm}<{\centering}p{4cm}}
			\hline
			\hline
			 & $E_{d_{z^2}}^{\text{bond}}$ (eV)   & $E_{d_{z^2}}^{\text{anti}}$ (eV)    & $E_{d_{x^2-y^2}}$ (eV)  & $E_{p_z}^{\text{apical O}}$ (eV) & $ \frac{E_{d_{z^2}}^{\text{bond}} -E_{d_{z^2}}^{\text{anti}}}{(E_{p_z}^{\text{apical O}} - E_{d_{z^2}}^{\text{bond}})(E_{d_{z^2}}^{\text{anti}} - E_{p_z}^{\text{apical O}}) }$ (eV$^{-1}$) & $V_{pd}$ (eV)    \\
			\hline

0 GPa                & -0.764                & 1.561                 & 0.760                  & 0.269                 & 1.742                 & 1.496                 \\
30 GPa                  & -0.730                 & 1.539                 & 0.805                 & 0.986                 & 2.391                & 1.804                 \\
100 GPa                 & -0.554                & 1.415                 & 0.951                 & 1.497                 & -11.710                & 2.186                 \\
Unstrained         & -0.764                & 1.561                 & 0.760                  & 0.269                 & 1.742                 & 1.496                 \\
-2\% strain          & -0.977                & 1.285                 & 0.884                 & 0.364                 & 1.832                & 1.558                 \\

			\hline
\hline	
\end{tabular}}\label{tableS2}
\vspace{3cm}
\end{table}

\begin{figure}[!h]
	\centering
	\includegraphics[scale=0.45,angle=0]{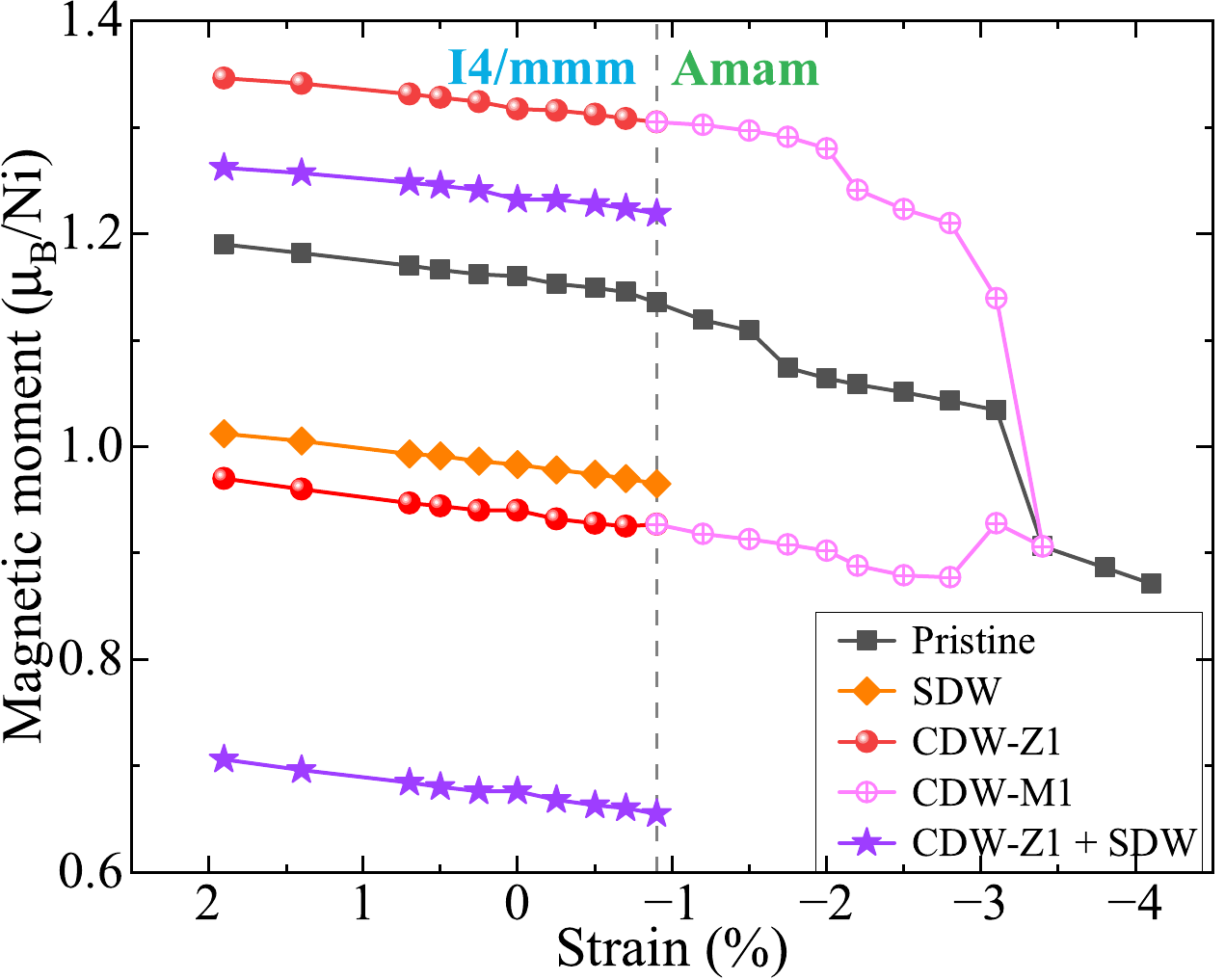}\\
	\caption{Strain-dependent magnetic moments of Ni atoms in different density wave orders. The structural distortion induced by CDW results in two distinct Ni sites with split magnetic moments.}\label{dw-magnetism}
\end{figure}

\begin{table}[!h]
	\caption{Calculated magnetic couplings ($J_1$-$J_5$) between Ni atoms for La$_3$Ni$_2$O$_{7}$ without strain and under -2\% compressive strain.}
	{\centering
		\begin{tabular}{lp{1cm}<{\centering}p{2cm}<{\centering}p{1.2cm}<{\centering}p{1.2cm}<{\centering}p{1.2cm}<{\centering}p{1.2cm}<{\centering}p{1.2cm}<{\centering}p{1.2cm}}
			\hline
			\hline
			& Strain &Space group &J$_1$ (eV)  &J$_2$ (eV)  &J$_3$ (eV)  &J$_4$ (eV)  &J$_5$ (eV)         \\
			\hline
			
			&\multicolumn{1}{r}{0\%} &$Amam$ &11.62 &7.86 &-2.91 &1.40 &-1.81\\
			&\multicolumn{1}{r}{-2\%} &$I4/mmm$ &18.34 &18.34 &-2.79 &-2.79 &5.99\\
			\hline
			\hline	
	\end{tabular}}\label{tableS3}
\end{table}

\end{document}